\documentclass[%
prb
,twocolumn%
, notitlepage
,superscriptaddress
,amsmath,amssymb,aps
, nofootinbib
]{revtex4-2}
\usepackage[dvipdfmx]{graphicx}
\usepackage{bm}
\usepackage{dcolumn}
\usepackage{color}
\usepackage{tabularx}
\usepackage{braket}
\usepackage{MnSymbol}
\usepackage{multirow}
\usepackage{xurl}

\allowdisplaybreaks

\begin{document}

\title{Physics-informed neural networks for solving functional renormalization group on a lattice}
\author{Takeru Yokota}
\email{takeru.yokota@riken.jp}
\affiliation{Interdisciplinary Theoretical and Mathematical Sciences Program (iTHEMS), RIKEN, Wako, Saitama 351-0198, Japan}
\date{\today}

\begin{abstract}
Addressing high-dimensional partial differential equations to derive effective actions within the functional renormalization group is formidable, especially when considering various field configurations, including inhomogeneous states, even on lattices. We leverage physics-informed neural networks (PINNs) as a state-of-the-art machine learning method for solving high-dimensional partial differential equations to overcome this challenge. In a zero-dimensional O($N$) model, we numerically demonstrate the construction of an effective action on an $N$-dimensional configuration space, extending up to $N=100$. Our results underscore the effectiveness of PINN approximation, even in scenarios lacking small parameters such as a small coupling.
\end{abstract}
\maketitle
\section{Introduction}
The utilization of the functional renormalization group (FRG) \cite{weg73,wil74,pol84,wet93} has gained widespread popularity as a non-perturbative theoretical tool across diverse fields, encompassing high-energy physics, condensed matter physics, and statistical physics (see Refs.~\cite{BERGES2002223, Gies2012, RevModPhys.84.299, DUPUIS20211} for comprehensive reviews). Central to the FRG is the utilization of functional differential equations (FDE), such as the Wetterich equation \cite{wet93}, which plays a pivotal role in describing the flow within the renormalization group (RG). Notably, the self-determination of the effective action (encompassing all the correlation information of a system) through the Wetterich equation enhances the precision and comprehensiveness of the FRG formalism. 
Despite these advantages, the absence of universal, efficient, and accurate algorithms for solving FDEs hampers an easily accessible and accurate determination of various properties. The quest for such algorithms or useful approximation schemes remains an open problem.

Power series expansions, such as the vertex expansion (the functional Taylor expansion) and the derivative expansion, are commonly employed in solving the FRG. In these approaches, the Wetterich equation transforms into an infinite hierarchy of differential equations for the expansion coefficients. These coefficients are subsequently truncated in a certain order to facilitate approximate solutions. However, the effective action $\Gamma[\varphi]$ is only valid for specific field configurations $\varphi(x)$, where $x$ is a spatio-temporal coordinate. For instance, the effective action derived from the vertex expansion is valid in the vicinity of the expansion point $\varphi(x)\approx \varphi_{\rm exp}(x)$, while that obtained from the derivative expansion is applicable when $\varphi(x)\approx \mathrm{const}$. In these cases, prior knowledge of the field's ground state is a prerequisite for calculations, limiting the ability to capture complex structures such as inhomogeneous states. Moreover, enhancing the accuracy of results often entails computationally demanding efforts in improving the truncation order.

The application of FRG extends beyond continuum models and is commonly employed in lattice models as well \cite{RevModPhys.84.299, DUPUIS20211}. On a finite lattice, the Wetterich equation becomes an $(N_{\rm DOF}+1)$-dimensional partial differential equation (PDE) involving $N_{\rm DOF}$ degrees of freedom for the field variables and the RG scale. While a finite-dimensional PDE might appear more amenable to numerical analysis than an FDE, the computational complexity of calculations with a large $N_{\rm DOF}$ grows exponentially when a computational grid is assigned to each field component. Therefore, even in lattice models, approximations based on power series expansions remain commonly employed.

This study aims to demonstrate that machine learning offers a novel framework for solving the FRG applied to lattice models with large $N_{\rm DOF}$ as an alternative to power series expansions. Among the array of recently developed machine learning methods for handling high-dimensional PDEs \cite{https://doi.org/10.1002/aic.690381003, https://doi.org/10.1002/cnm.1640100303, 712178, e_deep_2017, doi:10.1073/pnas.1718942115, doi:10.1142/9789811280306_0018, e_deep_2018, khoo_solving_2018, SIRIGNANO20181339, beck_machine_2019, RAISSI2019686, fuj19, NABIAN201914}, we leverage the physics-informed neural networks (PINNs) \cite{https://doi.org/10.1002/aic.690381003, 712178, RAISSI2019686}.
PINNs can be applied to various PDEs and involve optimizing a differentiable neural network (NN) to satisfy PDE and boundary conditions, providing a solution for a domain of input variables' space rather than a single point. Due to its grid-free characteristic, PINNs are particularly advantageous for handling high-dimensional inputs, as demonstrated in recent applications to high-dimensional PDEs \cite{GUO2022115523,he2023learning,cen2023deep, TANG2023111868,hu2023tackling,hu2023biasvariance,hu2023hutchinson}, including $10^5$-dimensional cases \cite{hu2023tackling,hu2023hutchinson}. In such scenarios, the limitations imposed by $N_{\rm DOF}$ are naturally expected to relax, implying the possibility of simultaneously constructing effective actions for various field configurations, including inhomogeneous states. Moreover, the universal approximation theorem \cite{cybenko_approximation_1989, HORNIK1989359,256500} suggests that NNs can serve as accurate approximations for effective action. 

In the subsequent sections, we present a PINN-based method for solving the FRG applied to a lattice (PINN-LFRG). We outline a methodology for representing the effective action using a differentiable NN, which is trained to satisfy the Wetterich equation. Furthermore, we provide numerical demonstrations of the scalability and accuracy of this approach in the zero-dimensional $\mathrm{O}(N)$ model. Here, PDEs with $N_{\rm DOF}+1=N+1\leq 101$ dimensions are solved within a few hours. The effective action and self-energy are computed across a domain of the field space simultaneously, exhibiting superior or comparable accuracies when contrasted with results obtained through perturbative and large-$N$ expansions, spanning various choices of coupling strength and $N$. Additionally, the $\mathrm{O}(N)$ symmetry for the effective action is successfully reproduced through training on the Wetterich equation. These findings underscore the feasibility of utilizing NNs to approximate the effective action even without a small parameter, such as a small coupling. We note that our purpose of solving FRG flow equations differs from that in a recent machine-learning-based FRG study \cite{PhysRevLett.129.136402}, which focuses on the dimensionality reduction of the four-point vertex function as given by FRG.

The remaining part of this paper is organized as follows: In the following section, we briefly summarize the FRG formulation for bosons in a lattice and illustrate our idea for applying PINNs to solve the Wetterich equation. In Sec.~\ref{sec: Numerical demonstration}, we present a numerical demonstration of our approach in the zero-dimensional $\mathrm{O}(N)$ model. Section \ref{sec: Conclusion} is devoted to the conclusion.

\section{General formulation \label{sec: General formulation}}
We focus on the FRG applied to bosons in a $d$-dimensional space-time lattice. The action is represented by $S(\boldsymbol{\varphi})$, where $\boldsymbol{\varphi}=\lbrace\varphi_{n,\alpha}\rbrace_{n,\alpha}$ is a real bosonic field. Here, the $d$-dimensional vector $n$ indicates a lattice site and $\alpha$ is the internal degrees of freedom index. The total degrees of freedom for this system are given by $N_{\rm DOF}=VN_{\rm IDOF}$, where $V$ denotes the lattice volume, and $N_{\rm IDOF}$ is the internal degree of freedom. The imaginary-time formalism is employed, and all quantities are expressed in lattice units.

We adhere to the formalism outlined by Wetterich \cite{wet93}. Following this formalism, a regulator term is introduced into the action to induce the RG flow:
\begin{align}
    S_k(\boldsymbol{\varphi})
    =
    S(\boldsymbol{\varphi})
    +
    \frac{1}{2}
    \sum_{n,\alpha,n,\alpha'}
    \varphi_{n,\alpha}R_{k,n-n'}^{\alpha\alpha'}\varphi_{n',\alpha'}.
\end{align}
The regulator $R_{k,n-n'}^{\alpha\alpha'}$ is a predefined function acting as an artificial mass, designed to dampen fluctuations with momenta smaller than the RG scale $k$. In the momentum space, the regulator must adhere to the following conditions:
\begin{subequations}
\begin{align}
    \label{eq:rcond_a}
    \lim_{p^2/k^2\to 0} \tilde{R}_k(p) &>0,
    \\
    \label{eq:rcond_b}
    \tilde{R}_{k_{\rm IR}\to 0}(p)&=0,
    \\
    \label{eq:rcond_c}
    \tilde{R}_{k_{\rm UV}\to \infty}(p)&=\infty.
\end{align}
\end{subequations}
For simplicity, we have omitted the indices for the internal degrees of freedom. The first condition signifies the suppression of infrared fluctuations, while the second condition ensures that all fluctuations are included at a small infrared scale $k_{\rm IR}$. The final condition is crucial for determining the initial condition of the RG flow. It ensures that the system becomes classical, described by $S_{k_{\rm UV}}(\boldsymbol{\varphi})$, at a large ultraviolet scale $k_{\rm UV}$. In terms of the path integral introduced below, this condition validates the saddle-point approximation at $k=k_{\rm UV}$. Compared to the continuous one, the distinctions in the lattice setup lie in the dispersion relation and the restriction of the momentum to the Brillouin zone. An appropriate regulator choice that accommodates these differences is discussed in Ref.~\cite{dupuis_non-perturbative_2008}.

With this regulator, one can define the effective average action $\Gamma_k(\boldsymbol{\varphi})$, which interpolates between the bare action $S(\boldsymbol{\varphi})$ and the effective action $\Gamma(\boldsymbol{\varphi})$. The definition is:
\begin{align}
    \label{eq:gammak_def}
    \Gamma_k(\boldsymbol{\varphi})
    =&
    \sup_{\boldsymbol{J}}
    \left(
    \sum_{n, \alpha}J_{n,\alpha}\varphi_{n,\alpha}-\ln Z_k(\boldsymbol{J})
    \right)
    \notag
    \\
    &
    -
    \frac{1}{2}
    \sum_{n,\alpha, n, \alpha'}
    \varphi_{n,\alpha}
    R_{k,n-n'}^{\alpha\alpha'}
    \varphi_{n',\alpha'},
\end{align}
with the path-integral form of the partition function
\begin{align}
    \label{eq:Zk}
    Z_k(\boldsymbol{J})
    =&
    \int d\boldsymbol{\varphi}
    e^{-S_k(\boldsymbol{\varphi})
    +
    a^d\sum_{n,\alpha}J_{n,\alpha}\varphi_{n,\alpha}}.
\end{align}
The condition $\lim_{k\to 0}\Gamma_{k}(\boldsymbol{\varphi})=\Gamma(\boldsymbol{\varphi})$ immediately follows from Eq.~\eqref{eq:rcond_b}. From the saddle-point approximation validated by Eq.~\eqref{eq:rcond_c}, we have $\Gamma_{k_{\rm UV}}(\boldsymbol{\varphi})=S(\boldsymbol{\varphi})+\mathrm{const}$. The RG flow equation is derived as an ($N_{\rm DOF}+1$)-dimensional PDE by the derivative of Eq.~\eqref{eq:gammak_def} with respect to $k$:
\begin{align}
    \label{eq:wet}
    \partial_k \Gamma_k(\boldsymbol{\varphi})
    =&
    \frac{1}{2}
    \mathrm{tr}
    \left[
    \partial_k R_k
    \left(
    \frac{\partial^2 \Gamma_k(\boldsymbol{\varphi})}{\partial \boldsymbol{\varphi} \partial \boldsymbol{\varphi}}
    +
    R_k
    \right)^{-1}
    \right],
\end{align}
which is known as the Wetterich equation. Here, the inverse is defined by
\begin{align}
    &\sum_{n'\alpha'}
    \left(
    \frac{\partial^2 \Gamma_k(\boldsymbol{\varphi})}{\partial \boldsymbol{\varphi} \partial \boldsymbol{\varphi}}
    +
    R_k
    \right)^{-1}_{n\alpha,n'\alpha'}
    \left(
    \frac{\partial^2 \Gamma_k[\varphi]}{\partial \varphi_{n',\alpha'} \partial \varphi_{n'',\alpha''}}
    +
    R_{k,n'-n''}^{\alpha'\alpha''}
    \right)
    \notag
    \\
    &=
    \delta_{n,n''}\delta_{\alpha\alpha''}.
\end{align}

In principle, Eq.~\eqref{eq:wet} determines $\Gamma(\boldsymbol{\varphi})$, encompassing all the thermodynamic properties and correlations. Typically, Taylor series expansions, including vertex and derivative expansions, are employed. These expansions yield an approximate calculation of $\Gamma(\boldsymbol{\varphi})$ for a specific configuration of $\varphi_{n,\alpha}$. However, there is currently no established method to accurately and efficiently obtain $\Gamma(\boldsymbol{\varphi})$ for a broad domain of the $\varphi_{n,\alpha}$ space. Our goal is to propose a promising candidate for such a method.

\subsection{PINNs for the Wetterich equation}
Initially, calculations involving large $N_{\rm DOF}$ may appear computationally challenging, given their complexity, which grows exponentially when a grid is associated with each component $\varphi_{n,\alpha}$. However, our approach is rooted in the resilience of PINNs to this issue, given its grid-free nature and applicability to high-dimensional PDEs \cite{GUO2022115523,he2023learning,cen2023deep, TANG2023111868,hu2023tackling,hu2023biasvariance,hu2023hutchinson}. In PINNs, the solution is represented by a differentiable NN, eliminating the need for discretization in numerical differentiation. The NN is optimized to satisfy the PDE and the boundary conditions (BCs) using backpropagation. The optimization function may take the form $L=L_{\rm PDE}+\lambda L_{\rm BC}$ \cite{RAISSI2019686}, where $L_{\rm PDE}$ ($L_{\rm BC}$) reaches its minimum if, and only if, the NN satisfies the PDE (BC) for any input, and $\lambda$ is a positive hyperparameter to adjust the relative scale of the two terms. 

The presence of both terms $L_{\rm PDE}$ and $L_{\rm BC}$ can pose challenges. For example, tuning $\lambda$ for efficient optimization convergence can be required. However, in our case of the initial value problem, $L_{\rm BC}$ can be omitted with an appropriate choice of the ansatz on $\Gamma_k(\boldsymbol{\varphi})$ similar to Ref.~\cite{712178}. We make such an ansatz based on the decomposition:
\begin{align}
    \Gamma_k(\boldsymbol{\varphi})
    =
    S(\boldsymbol{\varphi})
    +
    \Gamma_{\rm RG}(l,\boldsymbol{\varphi}),
\end{align}
where $l=\ln(k_{\rm UV}/k)$. Since the initial condition is $\Gamma_{k_{\rm UV}}(\boldsymbol{\varphi})
    =
    S(\boldsymbol{\varphi})$, the RG-induced part $\Gamma_{\rm RG}(l,\boldsymbol{\varphi})$ satisfies $\Gamma_{\rm RG}(0,\boldsymbol{\varphi})=0$. We further decompose $\Gamma_{\rm RG}(l,\boldsymbol{\varphi})$ as 
\begin{align}
    \Gamma_{\rm RG}(l,\boldsymbol{\varphi})
    =
    \gamma_{\rm free}(l)
    +
    \gamma(l,\boldsymbol{\varphi}).
\end{align}
Here, $\gamma_{\rm free}(l)$ represents the constant term originating from the free quadratic term $S_{\rm free}(\boldsymbol{\varphi})$ of $S(\boldsymbol{\varphi})$. In other words, it is the solution when $\Gamma_{k}(\boldsymbol{\varphi})$ on the right-hand side of Eq.~\eqref{eq:wet} is substituted by $S_{\rm free}(\boldsymbol{\varphi})$. 
The remaining term $\gamma(l,\boldsymbol{\varphi})$ constitutes the non-trivial interaction-induced part, corresponding to the shift in the free energy. By imposing $\gamma(0,\boldsymbol{\varphi})=0$, we replace $\gamma(l,\boldsymbol{\varphi})$ with an NN. A conceivable choice is:
\begin{align}
    \label{eq:gamma_nn}
    \gamma(l,\boldsymbol{\varphi})\approx\gamma(l,\boldsymbol{\varphi};\boldsymbol{\theta})=\mathrm{NN}_{\boldsymbol{\theta}}(l,\boldsymbol{\varphi})-\mathrm{NN}_{\boldsymbol{\theta}}(0,\boldsymbol{\varphi}),
\end{align}
where $\mathrm{NN}_{\boldsymbol{\theta}}(l,\boldsymbol{\varphi})$ is a differentiable NN with parameters $\boldsymbol{\theta}$.

A possible choice of $L=L_{\rm PDE}$ to train $\gamma(l,\boldsymbol{\varphi};\boldsymbol{\theta})$ is \begin{align}
    \label{eq:Loptim}
    L_{\boldsymbol{\theta}}
    &=
    \mathop{\mathbb{E}}_{
    \substack{\boldsymbol{\varphi}\sim \mathcal{P}_{\boldsymbol{\varphi}}
    \\
    l\sim \mathcal{P}_{l}}}
    \left[
    \left(
    \partial_l \Gamma_k^{\boldsymbol{\theta}}(\boldsymbol{\varphi})
    -
    \frac{1}{2}
    \mathrm{tr}
    \partial_l R_{k}
    \left(
    \frac{\partial^2 \Gamma_k^{\boldsymbol{\theta}}(\boldsymbol{\varphi})}{\partial \boldsymbol{\varphi} \partial \boldsymbol{\varphi}}
    +
    R_k
    \right)^{-1}
    \right)^2
    \right],
\end{align}
with $\Gamma_l^{\boldsymbol{\theta}}(\boldsymbol{\varphi})=S(\boldsymbol{\varphi})+\gamma_{\rm free}(l)+\gamma(l,\boldsymbol{\varphi};\boldsymbol{\theta})$, we introduce probability distributions $\mathcal{P}_{\boldsymbol{\varphi}}$ and $\mathcal{P}_{l}$, defined for the $\boldsymbol{\varphi}$-space and $l\in [0,l_{\rm end}]$, with $l_{\rm end}=\ln(k_{\rm UV}/k_{\rm IR})$, respectively. In practice, the expectation value is approximately evaluated using a finite number of collocation points $\lbrace (l^{(i)},\boldsymbol{\varphi}^{(i)}) \rbrace_{i=1}^{N_{\rm col}}$ sampled according to $\mathcal{P}_{\boldsymbol{\varphi}}$ and $\mathcal{P}_{l}$. Naively, if one is interested in a specific configuration $\boldsymbol{\varphi}=\boldsymbol{\varphi}_{\rm target}$, then $\mathcal{P}_{\boldsymbol{\varphi}}$ should be chosen as to sample the neighborhoods of $\boldsymbol{\varphi}_{\rm target}$ at high rates. A caveat is that, even in such a case, $\mathcal{P}_{\boldsymbol{\varphi}}$ should be sufficiently broad for learning $\boldsymbol{\varphi}$ derivatives, i.e., the $\boldsymbol{\varphi}$-dependence of $\gamma(l,\boldsymbol{\varphi};\boldsymbol{\theta})$. We surmise that the breadth should have the scale of the fluctuation $\sqrt{\braket{(\varphi_{n,\alpha}-\varphi_{{\rm target}, n,\alpha})^2}}$ for each direction $\varphi_{n,\alpha}$ to describe correlations.

The PINN-LFRG method described above is expected to offer advantages for complex structures, such as inhomogeneous states, over conventional FRG approximations. Specifically, the PINN-LFRG demonstrates improved scaling of computational complexity with respect to $N_{\mathrm{DOF}}$ compared to the vertex expansion, as illustrated in Appendix \ref{sec: Comparison of complexity}.

\section{Numerical demonstration in the zero-dimensional $\mathrm{O}(N)$ model \label{sec: Numerical demonstration}}
To illustrate how PINN-LFRG works, we apply it to the zero-dimensional $\mathrm{O}(N)$ model, which possesses an exact solution. The action is given by:
\begin{align}
    \label{eq:actionON}
    S(\boldsymbol{\varphi})=\frac{1}{2}m^2\boldsymbol{\varphi}^2+\frac{g}{4!}(\boldsymbol{\varphi}^2)^2,
\end{align}
where $\boldsymbol{\varphi}=(\varphi_1,\ldots, \varphi_N)$ represents an $N$-component scalar field, and $m$ and $g$ are the mass and coupling, respectively. 

This model gives the total degree of freedom by $N_{\rm DOF}=N_{\rm IDOF}=N$ due to $V=1$. We investigate the scalability with respect to $N_{\rm DOF}$ by increasing $N$.\footnote{The Wetterich equation for this model can be reduced to a two-dimensional PDE with variables $l$ and $\rho=\boldsymbol{\varphi}^2/2$, but we do not use this reduction to investigate scalability} We also assess accuracy by comparing the results of the interaction-induced effective action $\gamma(l,\boldsymbol{\varphi})$ and the RG-induced self-energy $\sigma_\alpha(l,\boldsymbol{\varphi})=\partial^2\gamma(l,\boldsymbol{\varphi})/\partial \varphi_\alpha^2$ to those from the exact calculation, perturbative expansion up to the leading order, and large-$N$ expansion up to O(1), which are summarized in Appendix \ref{app: Exact calculation}. Note that $\tilde{g}=Ng/m^4$ is the dimensionless control parameter determining the perturbative region as $\tilde{g}\ll 1$ due to $\langle \boldsymbol{\varphi}^2 \rangle\sim N/m^2$ \cite{Keitel_2012}. With the regulator $R_k^{\alpha\alpha'}=k_{\rm UV}^{2}e^{-2l}\delta_{\alpha\alpha'}$, our parameters satisfy $m^2/k_{\rm UV}^2=0.01$ and $\tilde{g} \ll 100$, which validate the stationary point approximation at $l=0$ and realize $\Gamma_{k_{\rm UV}}(\boldsymbol{\varphi})\approx S(\boldsymbol{\varphi})$. We set $l_{\rm end}=5$.

\begin{figure}[!t]
 \begin{center}
   \includegraphics[width=\columnwidth]{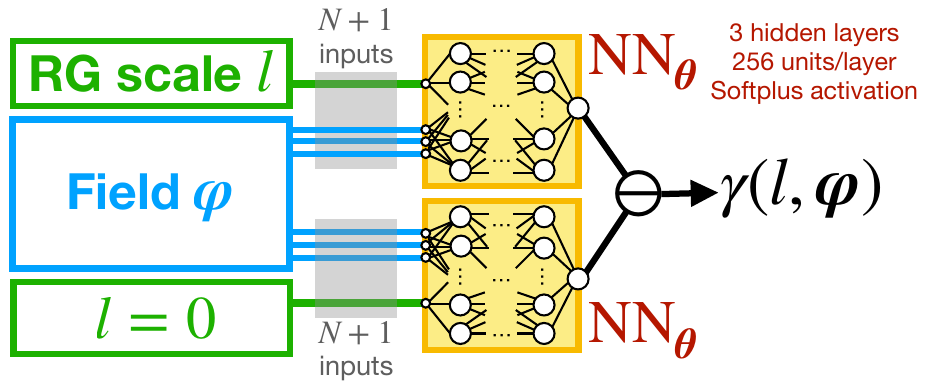}
   \caption{The schematic picture of our NN architecture for the interaction-induced effective action $\gamma(l,\boldsymbol{\varphi})$.\label{fig:schematic}}
 \end{center}
\end{figure}
Our ansatz on $\gamma(l,\boldsymbol{\varphi})$ is based on Eq.~\eqref{eq:gamma_nn}. Our $\mathrm{NN}_{\boldsymbol{\theta}}(l,\boldsymbol{\varphi})$ is a fully connected NN composed of 3 hidden layers with 256 units per layer and the differentiable softplus activation function. We find that this choice of NN shows successful convergence in the pretraining described below. Figure \ref{fig:schematic} depicts a schematic of our proposed NN architecture for $\gamma(l,\boldsymbol{\varphi})$. 

In our experience, the regularity of the matrix $\partial_{\boldsymbol{\varphi}}^2 \Gamma_k^{\boldsymbol{\theta}}(\boldsymbol{\varphi})+R_k$, which is needed for the matrix inverse in Eq.~\eqref{eq:Loptim}, is frequently broken for randomly chosen $\boldsymbol{\theta}$. We find that pretraining with some approximate analytical results remedies this problem. Specifically, we use the result of the first-order perturbation $\gamma^{\rm 1pt}(l,\boldsymbol{\varphi})$, employing the following optimization function:
\begin{align}
    \label{eq:Lpre}
    L_{\boldsymbol{\theta}}^{\rm pre}
    &=
    \mathop{\mathbb{E}}_{
    \substack{\boldsymbol{\varphi}\sim \mathcal{P}_{\boldsymbol{\varphi}}
    \\
    l\sim \mathcal{P}_{l}}}
    \left[\left(\gamma(l,\boldsymbol{\varphi};\boldsymbol{\theta})-\gamma^{\rm 1pt}(l,\boldsymbol{\varphi})\right)^2
    \right].
\end{align}
It should be noted that for stabilizing training on the Wetterich equation, the approximate solution doesn't always need high accuracy. In fact, when we use $\gamma^{\rm 1pt}(l,\boldsymbol{\varphi})$ during the pretraining phase, it significantly aids in successful training, even in non-perturbative cases, such as $\tilde{g}=10$.

The Adam optimizer \cite{kingma2017adam} is utilized to train the NN with Eqs.~\eqref{eq:Loptim} and \eqref{eq:Lpre}. All computations are executed on an NVIDIA A100 GPU with 40 GB of memory. Additional details about our training procedures are in Appendix \ref{app: Details about training}. The code for our numerical experiment is available at Ref.~\cite{yokota2024rep}.

We conducted computations for all the combinations of $N=1,10,100$ and $\tilde{g}=0.1,1,10$. In each case, the computational time for training is kept within 11 hours, ensuring the convergence of $L_{\boldsymbol{\theta}}$ and physical quantities; see Appendix \ref{app: Details about training} for details.

\begin{figure}[!t]
 \begin{center}
   \includegraphics[width=\columnwidth]{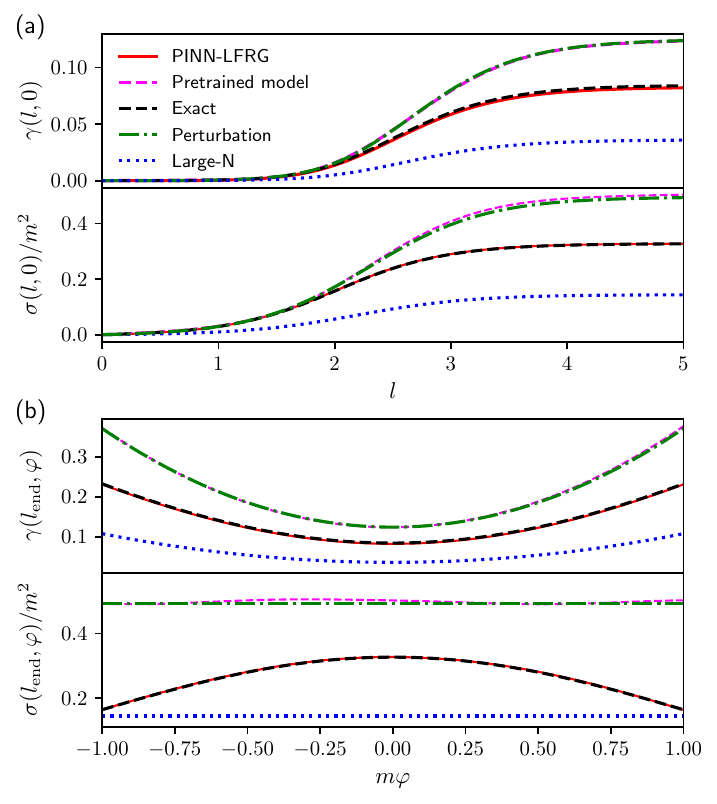}
   \caption{$\gamma(l,\varphi)$ and $\sigma(l,\varphi)$ in the case of $N=1$ and $\tilde{g}=1$. (a) $l$-dependence at $\varphi=0$ and (b) $\varphi$-dependence at $l=l_{\rm end}$. Results of the PINN-LFRG (red solid line), the pretrained model (magenta dashed line), exact calculation (black dashed line), perturbative expansion (green dotted--dashed line), and large-$N$ expansion (blue dotted line) are represented. \label{fig:N1}}
 \end{center}
\end{figure}
Figure~\ref{fig:N1} illustrates the results of $\gamma(l,\varphi)$ and $\sigma(l,\varphi)$ for $N=1$ and $\tilde{g}=1$. Specifically, the plot depicts the $l$-dependence at the vacuum expectation value $\varphi=0$ and the $\varphi$-dependence at $l=l_{\rm end}$. The results from exact calculations, perturbative and large-$N$ expansions, and the model after the pretraining are also presented. The perturbative and large-$N$ expansion results show considerable deviations from the exact ones since both $\tilde{g}=1$ and $1/N=1$ are not small. Notably, the training of the Wetterich equation successfully shifts values from those obtained by the perturbation approach toward the exact results. In all instances, our PINN-LFRG approach exhibits higher accuracy than the perturbative method and large-$N$ expansions. It is crucial to highlight that our approach provides solutions over a broad domain of $\varphi$, in contrast to the limitations of the vertex expansion method. 

\begin{figure}[!t]
 \begin{center}
   \includegraphics[width=\columnwidth]{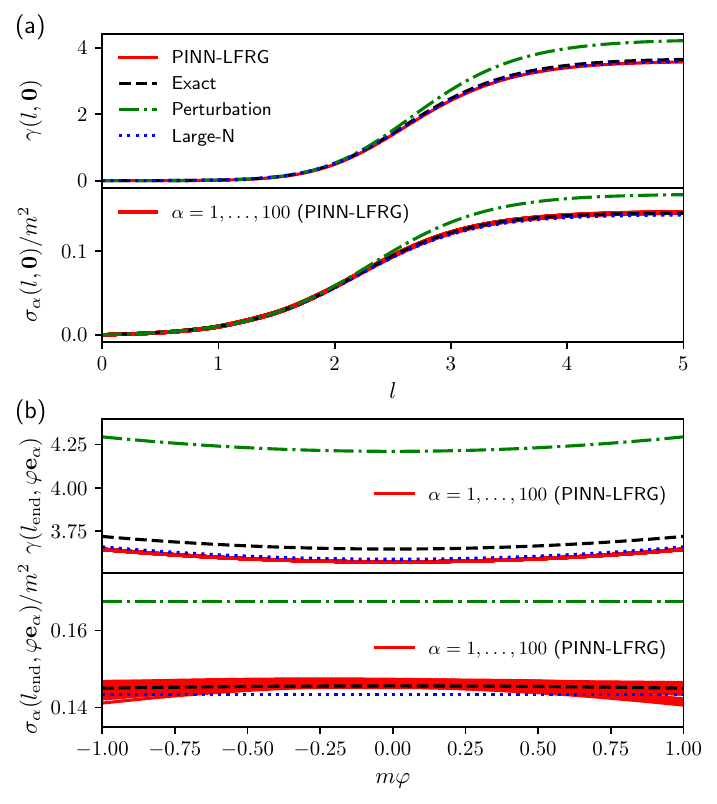}
   \caption{$\gamma(l,\varphi\boldsymbol{\mathrm{e}}_\alpha)$ and $\sigma_\alpha(l,\varphi\boldsymbol{\mathrm{e}}_\alpha)$ in the case of $N=100$ and $\tilde{g}=1$. (a) $l$-dependence at $\varphi=0$ and (b) $\varphi$-dependence at $l=l_{\rm end}$. The outcomes of PINN-LFRG are presented as the $N=100$ lines, except for $\gamma(l,\boldsymbol{0})$, where each line corresponds to a different choice of $\alpha$.
   \label{fig:N100}}
 \end{center}
\end{figure}
\begin{table*}[!t]
\centering
\caption{Relative errors of $\gamma=\gamma(l_{\rm end},\boldsymbol{0})$ and $\sigma=\sigma(l_{\rm end},\boldsymbol{0})$ compared to the exact values in percentage. The results of the perturbation, large-$N$ expansion, and PINN-LFRG are displayed. For PINN-LFRG, we present the ambiguity of the relative error of $\sigma$ estimated from the standard derivation (denoted by $\Delta \sigma$). The minus sign indicates underestimation. The best values in each column are highlighted with bold font.\label{tab:relerror}}
\newcolumntype{C}{>{\centering\arraybackslash}X}
\begin{tabularx}{2\columnwidth}{c|c|CCC|CCC|CCC}
    \hline\hline
    \multicolumn{2}{c|}{$N$} &\multicolumn{3}{c|}{1} & \multicolumn{3}{c|}{10} & \multicolumn{3}{c}{100} \\
    \multicolumn{2}{c|}{$\tilde{g}$} 
    & 0.1 & 1 & 10 & 0.1 & 1 & 10 & 0.1 & 1 & 10 \\
    \hline
    \hline
    \multirow{2}{*}{Perturb. (\%)} & $\gamma$
    & 6.2 & 47 & 275 & 2.1 & 19 & 129 & \textbf{1.7} & 15 & 110
    \\
    & $\sigma$
    & 7.6 & 51 & 228 & 2.3 & 19 & 109 & 1.7 & 15 & 92
    \\
    \hline
    \multirow{2}{*}{Large-$N$ (\%)} & $\gamma$
    & -65 & -57 & -40 & -16 & -14 & -8.4 & -1.9 & \textbf{-1.6} & \textbf{-0.95}
    \\
    & $\sigma$
    & -65 & -56 & -42 & -16 & -13 & -8.2 & -1.9 & -1.5 & -0.89
    \\
    \hline
    \multirow{3}{*}{PINN-LFRG (\%)} & $\gamma$
    & \textbf{-2.0} & \textbf{-2.2} & \textbf{-2.8} &  \textbf{-1.9} & \textbf{-2.1} & \textbf{-2.3} & -1.9 & -2.0 & -2.3
    \\
    & $\sigma$
    & \textbf{-0.17} & \textbf{0.12} & \textbf{0.76} & \textbf{0.16} & \textbf{0.46} & \textbf{0.42} & \textbf{-0.011} & \textbf{0.44} & \textbf{0.50}
    \\
    & $\Delta \sigma$
    & 0 & 0 & 0 & 0.27 & 0.18 & 0.24 & 0.38 & 0.29 & 0.26
    \\
    \hline\hline
\end{tabularx}
\end{table*}
\begin{figure}[!t]
 \begin{center}
   \includegraphics[width=\columnwidth]{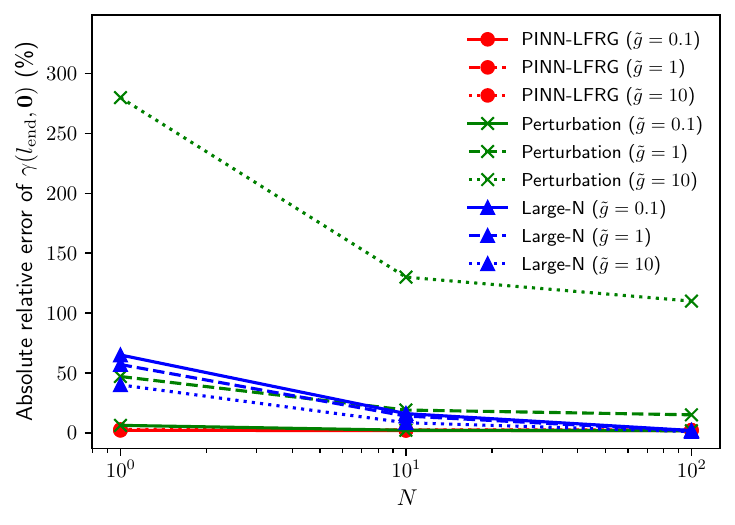}
   \caption{Absolute values of the relative errors of $\gamma=\gamma(l_{\rm end},\boldsymbol{0})$ compared to the exact values as a function of $N$. All the cases of PINN-LFRG, perturbation, and large-$N$ expansion with $\tilde{g}=0.1,1,10$ are depicted. \label{fig:err}}
 \end{center}
\end{figure}
Figure~\ref{fig:N100} illustrates the result for $N=100$ and $\tilde{g}=1$. With the exception of $\gamma(l,\boldsymbol{0})$, our results are presented as the $N=100$ lines corresponding to the $N$ directions in the $\boldsymbol{\varphi}$-space. This includes all the $\alpha=1,\ldots,100$ cases of $\gamma(l,\varphi \boldsymbol{\mathrm{e}}_\alpha)$ and $\sigma_\alpha(l,\varphi \boldsymbol{\mathrm{e}}_\alpha)$, where $\boldsymbol{\mathrm{e}}_\alpha$ denotes the unit vector in the $\varphi_\alpha$ direction. The PINN-LFRG results for different $\alpha$ closely match, with differences being imperceptible in $\gamma(l_{\rm end},\varphi \boldsymbol{\mathrm{e}}_\alpha)$ and $\sigma_\alpha(l,\boldsymbol{0})/m^2$. Even in $\sigma_\alpha(l_{\rm end},\varphi \boldsymbol{\mathrm{e}}_\alpha)$, all the results from our approach are as close to the exact result as those of the large-$N$ expansion, which is expected to be accurate for $N=100$. These findings indicate that the NN automatically captures the $\mathrm{O}(N)$ symmetry, enabling a simultaneously accurate solution for a domain of the high-dimensional configuration space. 

Table~\ref{tab:relerror} summarizes the relative errors of $\gamma(l_{\rm end},\boldsymbol{0})$ and $\sigma(l_{\rm end},\boldsymbol{0})$ compared to the exact values for all values of $N$ and $\tilde{g}$. In the case of PINN-LFRG for $N>1$, we determine $\sigma(l_{\rm end},\boldsymbol{0})$ by averaging $\sigma_\alpha(l_{\rm end},\boldsymbol{0})$ with respect to $\alpha=1,\ldots, N$, and we derive the standard deviation. To show the tendency, we plot the absolute values of the relative errors of $\gamma(l_{\rm end},\boldsymbol{0})$ compared to the exact values as a function of $N$ in Fig.~\ref{fig:err}; almost the same tendency is seen for $\sigma(l_{\rm end},\boldsymbol{0})$. For all $N$ and $\tilde{g}$ values, the errors of PINN-LFRG are within 3\% for $\gamma(l_{\rm end},\boldsymbol{0})$ and 1\% for $\sigma(l_{\rm end},\boldsymbol{0})$ even if the standard deviations are taken into account. Particularly, PINN-LFRG is accurate even for the non-perturbative and small-$N$ regions, where both the perturbative and large-$N$ expansions break down. This suggests that the NN is a promising tool for providing accurate approximation independently of the existence of a small parameter.

\section{Conclusion \label{sec: Conclusion}}
This study introduces PINN-LFRG as a novel framework for solving the Wetterich equation on a finite lattice. The approach demonstrates the ability to simultaneously derive an effective action for various field configurations. The proposed procedure involves representing the effective action through an NN and optimizing it. The demonstration in the zero-dimensional $\mathrm{O}(N)$ model indicates the feasibility of calculations involving a substantial number of degrees of freedom, around $10^2$ or more, with NNs effectively approximating the effective action without the reliance on a small parameter. 

Our analysis can be readily extended to models incorporating temporal and spatial degrees of freedom. An intriguing avenue for further exploration is the investigation of inhomogeneous states in scalar models, such as solitons, within our framework, building upon existing work on this topic \cite{Bolognesi:2016zjp, Nitta:2017uog, Gorsky:2018lnd}. Extending the approach to fermionic systems poses a substantial challenge since there is currently no efficient method for constructing NNs for Grassmann variables. However, one could apply our approach to fermionic systems by introducing bosonic auxiliary fields, for example. An exciting application in this direction is the adaptation of our method to density functional theory \cite{PhysRev.136.B864, PhysRev.140.A1133, RevModPhys.71.1253}, a standard tool for analyzing many-body systems. This has been extended to apply to lattice models, such as the Hubbard model \cite{Carrascal_2015}. We anticipate that our approach holds promise for the FRG-based formalism of density functional theory, a framework that has seen recent developments \cite{pol02,sch04,kem13,ren15,kem17a,lia18,yok18,yok18b,yok19,yok20,yok21,yok21b,yok22}. 

\begin{acknowledgments}
The author thanks Taiki Miyagawa for carefully reading the manuscript and making valuable comments. The author also thanks Gergely Fej\H{o}s, Tetsuo Hatsuda, Tomoya Naito, Osamu Sugino, and Junji Yamamoto for valuable discussions. The RIKEN Special Postdoctoral Researchers Program supported the author.
\end{acknowledgments}

\appendix

\begin{widetext}
\section{Comparison of complexity \label{sec: Comparison of complexity}}
As delineated in the main text, the PINN-LFRG approach is anticipated to offer advantages for analyzing complex structures, such as inhomogeneous states, when compared to conventional methods, including the vertex expansion. We discuss this advantage from the perspective of computational complexity. In the context of the lattice setup described in the main text, we concentrate on the scaling relative to $N_{\mathrm{DOF}}$.

The derivative expansion is ineffective for systems with large field gradients, which may be described by the vertex expansion, albeit at a significantly greater computational cost than in homogeneous cases. The computational complexity for the $i$th-order vertex expansion scales as $O(N_{\mathrm{DOF}}^{i+2})$ or $O(N_{\mathrm{DOF}}^{i+3})$, derived as follows: The $i$th-order vertex expansion around an inhomogeneous field profile $\boldsymbol{\varphi} = \boldsymbol{\varphi}_{\mathrm{inhom}}$ produces the flow equations for 
\begin{align}
    \Gamma^{(j)}_{k;n_1,\ldots,n_j}
    &=
    \frac{\partial^{j} \Gamma_k (\boldsymbol{\varphi}_{\mathrm{inhom}})}{\partial \varphi_{n_1}\cdots \partial \varphi_{n_j}}\quad (j=0,\ldots, i),
\end{align}
where $n_1,\ldots, n_j$ are site indices. For simplicity, we have omitted the internal degrees of freedom introduced in the main text. The computational bottleneck is the calculation of $\Gamma^{(i)}_{k;n_1,\ldots,n_i}$, governed by the following flow equation:
\begin{align}
    \partial_k \Gamma_{k;n_1,\ldots,n_{i}}^{(i)}
    &=
    \frac{1}{2}
    \mathrm{tr}
    \left[
    \text{matrix products of $\partial_k R_k$, $[\Gamma_k^{(2)}+R_k]^{-1}$, and $[\Gamma_{k;m_1,\ldots,m_{j-2}}^{(j)}]$ ($j=3,\ldots,i+2$)}
    \right],
\end{align}    
where $[\Gamma_{k;m_1,\ldots,m_{j-2}}^{(j)}]$ is an $N_{\mathrm{DOF}}\times N_{\mathrm{DOF}}$ matrix defined by
\begin{align}
    [\Gamma_{k;m_1,\ldots,m_{j-2}}^{(j)}]_{n,n'}
    &=
    \Gamma^{(j)}_{k;m_1,\ldots,m_{j-2},n,n'}
\end{align}
and each $m_l$ ($l=1,\ldots,j-2$) represents any of $n_{1},n_{2},\ldots,n_{i}$. The computational complexity of evaluating the trace is $O(N_{\mathrm{DOF}}^2)$ or $O(N_{\mathrm{DOF}}^3)$ depending on the algorithm.\footnote{The complexity scales as $O(N_{\mathrm{DOF}}^3)$ for direct evaluation of the matrix product and can be reduced to $O(N_{\mathrm{DOF}}^2)$ using methods such as the Hutchinson trace estimator \cite{doi:10.1080/03610919008812866}.} Since the flow equation is calculated for all combinations of $(n_1,\ldots,n_i)$, the complexity escalates to $O(N_{\mathrm{DOF}}^{i+2})$ or $O(N_{\mathrm{DOF}}^{i+3})$. In contrast, in homogeneous cases, the complexity is only $O(N_{\mathrm{DOF}}^{i})$, as the number of independent components of $\Gamma_{k;n_1,\ldots,n_{i}}^{(i)}$ is $O(N_{\mathrm{DOF}}^{i-1})$ and the trace evaluation requires merely $O(N_{\mathrm{DOF}})$ due to translational symmetry.

The complexity of the PINN-LFRG model is assessed as either $O(N_{\mathrm{DOF}}^2)$ or $O(N_{\mathrm{DOF}}^3)$, depending on the specific algorithm employed. This complexity arises primarily due to the evaluation of the trace in Eq.~\eqref{eq:Loptim}. These findings indicate that for large values of $N_{\mathrm{DOF}}$, the computational demands of the PINN-LFRG are potentially less than those of the vertex expansion, even at the second order, where the complexity reaches $O(N_{\mathrm{DOF}}^4)$ or $O(N_{\mathrm{DOF}}^5)$. For a more comprehensive comparison, further discussion of the influence of additional factors, such as network size, on the total numerical effort would be desirable.

\section{Exact calculation, perturbative expansion, and large-$N$ expansion in the zero-dimensional $\mathrm{O}(N)$ model \label{app: Exact calculation}}
We summarize the numerical procedure for the exact calculation and the results of the perturbative and large-$N$ expansions for the interaction-induced effective action $\gamma(l,\boldsymbol{\varphi})=\gamma(l,\varphi)$ and the RG-induced self-energy $\sigma(l,\varphi)=\partial_\varphi^2 \gamma(l,\varphi)$ in the zero-dimensional $\mathrm{O}(N)$ model. We use the form of the regulator $R_k^{\alpha\alpha'}=r_k\delta_{\alpha\alpha'}$ as in the main text.

The exact results are obtained by directly evaluating the path integral of the partition function:
\begin{align}
    Z_l(\boldsymbol{J})
    =&
    \int d\boldsymbol{\varphi}
    e^{-\frac{1}{2}m_l^2\boldsymbol{\varphi}^2-\frac{g}{4!}(\boldsymbol{\varphi}^2)^2+\boldsymbol{J}\cdot\boldsymbol{\varphi}},
\end{align}
where $m_l^2=m^2+r_k$ represents the regulated mass squared. Due to the presence of an $\mathrm{O}(N-1)$ symmetry in the $\boldsymbol{\varphi}$-space perpendicular to $\boldsymbol{J}$, the integral can be simplified as follows:
\begin{align}
    \label{eq:ZkON}
    Z_l(J)
    =&
    \Omega_{N-1}
    \int_{-\infty}^\infty d\varphi
    e^{-\frac{1}{2}m_l^2\varphi^2-\frac{g}{4!}\varphi^4+J\varphi}
    Q_{N-2,l}(\varphi^2),
    \\
    Q_{N-2,l}(\varphi^2)
    =&
    \int_0^{\infty}dx x^{N-2} e^{-\frac{1}{2}\left(m_l^2+\frac{g}{6}\varphi^2\right)x^2-\frac{g}{4!}x^4}
    =
    \frac{1}{2}
    \left(\frac{6}{g}\right)^{\frac{N-1}{4}}
    \Gamma\left(\frac{N-1}{2}\right)
    U
    \left(\frac{N-1}{4},\frac{1}{2},\frac{3}{2g}\left(m_l^2+\frac{g}{6}\varphi^2\right)^2
    \right),
\end{align}
where we have introduced $J=\|\boldsymbol{J}\|$, the surface area of the unit $(N-1)$-sphere $\Omega_{N}=2\pi^{N/2}/\Gamma(N/2)$, the gamma function $\Gamma(x)$, and the Tricomi's confluent hypergeometric function $U(a,b,z)$. Let $J=J_{{\rm sup}, l}(\varphi)$ be an external field realizing $\braket{\varphi}=\varphi$, i.e., the solution of
\begin{align}
    \label{eq:Jphieq}
    \varphi
    =
    \frac{\partial \ln Z_l}{\partial J}(J_{{\rm sup}, l}(\varphi))
    =
    \frac{
    \int_{-\infty}^\infty dx\,
    x
    e^{-\frac{1}{2}m_l^2x^2-\frac{g}{4!}x^4+J_{{\rm sup},l}(\varphi)x}
    Q_{N-2,l}(x^2)
    }{
    \int_{-\infty}^\infty dx\,
    e^{-\frac{1}{2}m_l^2x^2-\frac{g}{4!}x^4+J_{{\rm sup},l}(\varphi)x}
    Q_{N-2,l}(x^2)
    }.
\end{align}
With this external field, the effective action and the self-energy are given by
\begin{align}
    \Gamma(l,\varphi)
    =&
    J_{{\rm sup},k}(\varphi)\varphi
    -
    \ln Z_k(J_{{\rm sup},k}(\varphi))
    -
    \frac{1}{2}r_k\varphi^2,
    \\
    \Sigma(l,\varphi)
    =&
    \partial_\varphi^2 \Gamma(l,\varphi)-m^2
    =
    \frac{1}{\Braket{\varphi^2}-\varphi^2}-m_l^2,
\end{align}
where correlation function $\Braket{\varphi^2}-\varphi^2$ is evaluated by
\begin{align}
    \label{eq:corrint}
    \Braket{\varphi^2}-\varphi^2
    =&
    \frac{\partial^2 \ln Z_l}{\partial J^2}
    (J_{{\rm sup},l}(\varphi))
    =
    \frac{
    \int_{-\infty}^\infty dx
    e^{-\frac{1}{2}m_l^2x^2-\frac{g}{4!}x^4+J_{{\rm sup},l}x}
    (x-\varphi)^2
    Q_{N-2,l}(x^2)
    }{
    \int_{-\infty}^\infty dx
    e^{-\frac{1}{2}m_l^2x^2-\frac{g}{4!}x^4+J_{{\rm sup},l}x}
    Q_{N-2,l}(x^2)
    }.
\end{align}
With these $\Gamma(l,\varphi)$ and $\Sigma(l,\varphi)$, we obtain
\begin{align}
    \label{eq:gam_Gam}
    \gamma(l,\varphi)
    =&
    \Gamma(l, \varphi)-\Gamma(0, \varphi)-\gamma_{\rm free}(l),
    \\
    \label{eq:sigma_Gam}
    \sigma(l,\varphi)
    =&
    \Sigma(l,\varphi)-\Sigma(0,\varphi),
\end{align}
where $\gamma_{\rm free}(l)=(N/2)\ln(m_l^2/m_0^2)$ is the solution of
\begin{align}
    \label{eq:free_flow_eq}
    \partial_l \gamma_{\rm free}(l)
    =
    \frac{1}{2}
    \sum_{\alpha=1}^{N}
    \partial_l r_k
    \left(
    \frac{\partial^2 S_{\rm free}(\boldsymbol{\varphi})}{\partial \boldsymbol{\varphi} \partial \boldsymbol{\varphi}}
    +
    r_k
    \right)^{-1}_{\alpha\alpha},\quad
    \gamma_{\rm free}(0)=0,
\end{align}
with $S_{\rm free}(\boldsymbol{\varphi})=m^2 \boldsymbol{\varphi}^2/2$. We numerically solve Eq.~\eqref{eq:Jphieq} for $J_{{\rm sup},l}(\varphi)$ by use of \texttt{scipy.optimize.fsolve} in SciPy. With this $J_{{\rm sup},l}(\varphi)$, we numerically evaluate Eqs.~\eqref{eq:ZkON} and \eqref{eq:corrint} to obtain $\gamma(l,\varphi)$ and $\sigma(l,\varphi)$. The integrals in Eqs.~\eqref{eq:ZkON}, \eqref{eq:Jphieq}, and \eqref{eq:corrint} are evaluated using the Gauss quadrature method implemented as \texttt{scipy.integrate.quad} in SciPy.

The perturbative and large-$N$ expansion results are obtained from Ref.~\cite{Keitel_2012}. By substituting the regulated mass squared $m_l^2$ into these expressions, the results at the scale $l$ for the effective action and self-energy up to the leading order are given by:
\begin{align}
    \Gamma(l, \varphi)
    =&
    S(\boldsymbol{\varphi})
    +
    N
    \frac{1+2N^{-1}}{24}
    \tilde{g}_l
    +
    \frac{1+2N^{-1}}{12} 
    \tilde{g}_l
    m_l^2\varphi^2
    +
    \gamma_{\rm free}(l)
    +
    O(\tilde{g}_l^2)
    ,
    \\
    \Sigma(l, \varphi)
    =&
    \partial_{\varphi}^2
    S(\boldsymbol{\varphi})
    -
    m^2
    +
    \frac{1+2N^{-1}}{6} 
    \tilde{g}_l m_l^2
    +
    O(\tilde{g}_l^2).
\end{align}
Here, the dimensionless quantity $\tilde{g}_l=Ng/m_l^4$ is employed as the expansion parameter instead of $g$. The result of the large-$N$ expansion up to $O(1)$ is expressed as follows:
\begin{align}
    \Gamma(l, \varphi)
    =&
    S(\boldsymbol{\varphi})
    +
    N\left(\frac{z_l-1}{4}-\frac{1}{2}\ln z_l \right)
    +
    \frac{1}{2}
    \ln \left(2-z_l\right)
    +
    \frac{1}{2}\left(z_l^{-1}-1\right)m_l^2\varphi^2
    +
    \gamma_{\rm free}(l)
    +
    O(1/N),
    \\
    \Sigma(l, \varphi)
    =&
    \partial_{\varphi}^2
    S(\boldsymbol{\varphi})
    -
    m^2
    +
    \left(z_l^{-1}-1\right)m_l^2
    +
    O(1/N),
\end{align}
with
\begin{align}
    z_l
    =&
    \frac{2}{1+\sqrt{1+\frac{2}{3}\tilde{g}_l}}.
\end{align}
With these $\Gamma(l, \varphi)$ and $\Sigma(l,\varphi)$, we obtain $\gamma(l,\varphi)$ and $\sigma(l,\varphi)$ from Eqs.~\eqref{eq:gam_Gam} and \eqref{eq:sigma_Gam}.

\section{Details about training \label{app: Details about training}}

\begin{table}[!b]
\centering
\caption{Computational times for the pretraining and the training of the Wetterich equation (labeled by Wetterich) on an NVIDIA A100 GPU. The results are consistent for $\tilde{g}=0.1,1,10$. \label{tab:ct}}
\begin{tabular}{c|ccc}
    \hline\hline
    $N$ &1 & 10 & 100 
    \\
    \hline
    Pretraining & 4m & 4m & 6m
    \\
    Wetterich & 6h & 7h & 11h
    \\
    \hline\hline
\end{tabular}
\end{table}
\begin{figure}[!b]
 \begin{center}
   \includegraphics[width=0.95\columnwidth]{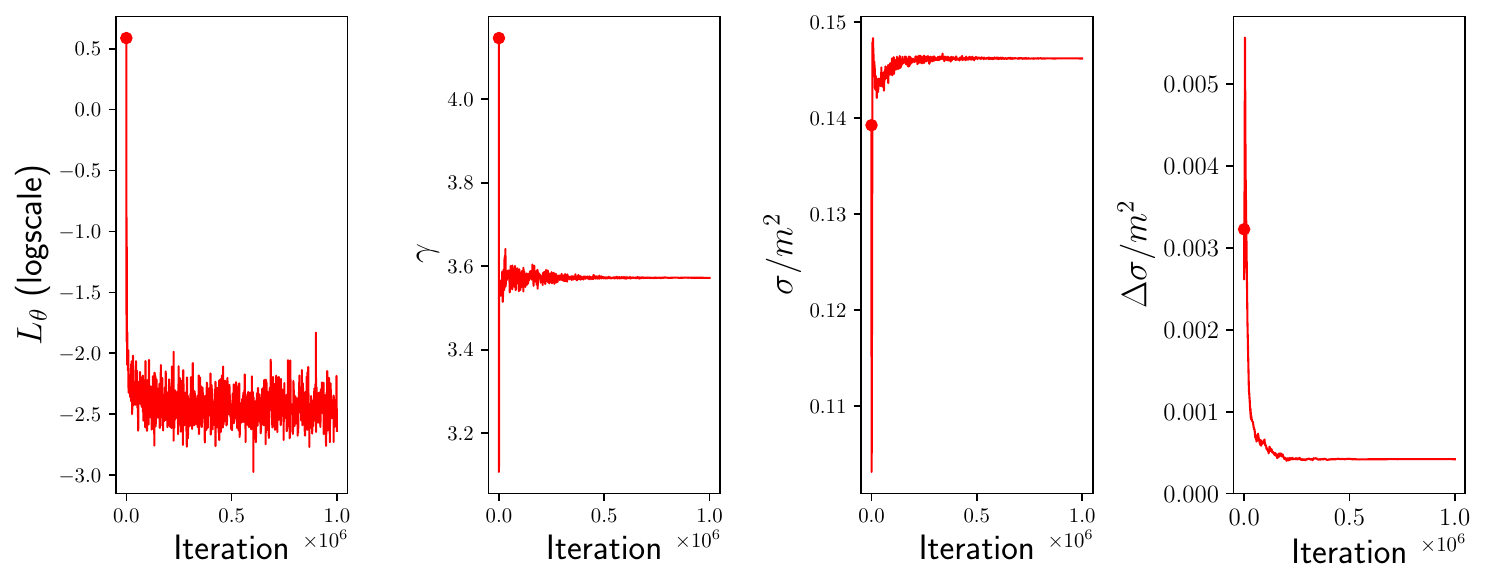}
   \caption{Learning curve and the histories of $\gamma=\gamma(l_{\rm end},\boldsymbol{0})$, the average $\sigma/m^2$, and the standard deviation $\Delta\sigma/m^2$ of $\sigma_\alpha(l,\boldsymbol{0})/m^2$ with respect to $\alpha$ in the case of $N=100$ and $\tilde{g}=1$. The red dots indicate the initial values. \label{fig:learning}}
 \end{center}
\end{figure}

We provide some details about the training and information about the convergence of our results. We optimize our NN to minimize
$L_{\boldsymbol{\theta}}$ ($L_{\boldsymbol{\theta}}^{\rm pre}$) in the main text for the training of the Wetterich equation (the pretraining). The expectations of these equations are approximately evaluated on a finite number of collocation points:
\begin{align}
    L_{\boldsymbol{\theta}}
    &\approx
    \frac{1}{N_{\rm col}}
    \sum_{n=1}^{N_{\rm col}}
    \left[
    \left(
    \partial_l \gamma(l^{(n)},\boldsymbol{\varphi}^{(n)};\boldsymbol{\theta})
    +
    \partial_l \gamma_{\rm free}(l^{(n)})
    -
    \frac{1}{2}
    \sum_{\alpha,\alpha'}
    \partial_l R_{k^{(n)}}^{\alpha\alpha'}
    \left(
    \frac{\partial^2 S(\boldsymbol{\varphi}^{(n)})}{\partial \boldsymbol{\varphi} \partial \boldsymbol{\varphi}}
    +
    \frac{\partial^2 \gamma(l^{(n)},\boldsymbol{\varphi}^{(n)};\boldsymbol{\theta})}{\partial \boldsymbol{\varphi} \partial \boldsymbol{\varphi}}
    +
    R_{k^{(n)}}
    \right)^{-1}_{\alpha',\alpha}
    \right)^2
    \right],
    \\
    L_{\boldsymbol{\theta}}^{\rm pre}
    &\approx
    \frac{1}{N_{\rm col}}
    \sum_{n=1}^{N_{\rm col}}
    \left[\left(\gamma(l^{(n)},\boldsymbol{\varphi}^{(n)};\boldsymbol{\theta})-\gamma^{\rm 1pt}(l^{(n)},\boldsymbol{\varphi}^{(n)})\right)^2
    \right],
\end{align}
where $l^{(n)}$($=\ln (k_{\rm UV}/k^{(n)})$) and $\boldsymbol{\varphi}^{(n)}$ are randomly sampled following the probability distributions $\mathcal{P}_{\boldsymbol{\varphi}}$ and $\mathcal{P}_{l}$. For $\mathcal{P}_{l}$, we adopt a uniform distribution within the interval $[0, l_{\rm end}]$. To sample the neighborhoods of $\boldsymbol{\varphi}=\boldsymbol{0}$, representing the vacuum expectation value, we define $\mathcal{P}_{\boldsymbol{\varphi}}$ such that the direction $\hat{\boldsymbol{n}}=\boldsymbol{\varphi}/\|\boldsymbol{\varphi}\|$ is uniformly sampled. The norm $\|\boldsymbol{\varphi}\|$ is sampled following a normal distribution $\mathcal{N}(0, N/m^2)$ without the sign, where the variance $N/m^2$ corresponds to the order of $\braket{\boldsymbol{\varphi}^2}$. It is noteworthy that the efficiently sampling neighborhoods of $\boldsymbol{\varphi}=\boldsymbol{0}$ for large $N$ is challenging if $\mathcal{P}_{\boldsymbol{\varphi}}$ is set to an $N$-dimensional normal distribution $\mathcal{N}(\boldsymbol{0}, m^{-2}\boldsymbol{1})$ or a uniform distribution in an $N$-dimensional box due to the curse of dimensionality. Specifically, we choose $N_{\rm col}=500$ collocation points, which are refreshed each time the optimization functions are assessed. 

For the numerical implementation, we employ Pytorch. The learning rate for the Adam optimizer is initially set to $10^{-4}$ and exponentially decays with a factor of 0.99999. The learning rate is fixed at $10^{-3}$ in the pretraining phase. The Xavier initialization \cite{pmlr-v9-glorot10a} is used for this pretraining. It is worth noting that the computational cost of evaluating the matrix inverse in $L_{\boldsymbol{\theta}}$ is substantial. To facilitate implementation, we directly compute the inversion using the \texttt{torch.linalg.inv} function in Pytorch. Efficiency enhancement, potentially utilizing alternative algorithms such as the Hutchinson trace estimator \cite{doi:10.1080/03610919008812866}, is reserved for future study. 

The training process involves $10^6$ iterations for the Wetterich equation and $10^5$ for the pretraining. Table \ref{tab:ct} provides an overview of the computational time required. With this iteration count, we observe the convergence of $L_{\boldsymbol{\theta}}$ and physical quantities. As illustrated in Fig.~\ref{fig:learning}, we present a learning curve along with the histories of $\gamma=\gamma(l_{\rm end},\boldsymbol{0})$ and the average $\sigma/m^2$, as well as the standard deviation $\Delta\sigma/m^2$ of $\sigma_\alpha(l,\boldsymbol{0})/m^2$ with respect to $\alpha$ for the case of $N=100$ and $\tilde{g}=1$. In the initial iterations, $L_{\boldsymbol{\theta}}$ rapidly decreases, and the physical quantities approach converges quickly. Subsequently, as the learning rate decays, physical quantities gradually converge. The diminishing $\Delta \sigma/m^2$ over iterations indicates successfully reproducing the $\mathrm{O}(N)$ symmetry during training.

\begin{table}[!t]
\centering
\caption{Relative errors of $\gamma(l_{\mathrm{end}},0)$ compared to the exact solutions for different numbers of the hidden layers and the units per layer with $N=1$ and $\tilde{g}=1$. The minus sign indicates underestimation. \label{tab: layer unit gamma}}
\newcolumntype{C}{>{\centering\arraybackslash}X}
\begin{tabularx}{0.5\columnwidth}{cc|CCC}
    \hline\hline
    \multirow{2}{*}{Relative error of $\gamma$ (\%)} & & \multicolumn{3}{c}{Number of units per layer}
    \\
    & & 64 & 128 & 256
    \\
    \hline
    \multirow{3}{*}{Number of hidden layers} 
    & 1 & -3.2 & -0.23 & -3.7
    \\
     & 2 & -2.2 & -2.2 & -2.2
    \\
     & 3 & -2.2 & -2.2 & -2.2
    \\
    \hline\hline
\end{tabularx}
\end{table}

\begin{table}[!t]
\centering
\caption{Relative errors of $\sigma=\sigma(l_{\mathrm{end}},0)$ compared to the exact solutions for different numbers of the hidden layers and the units per layer with $N=1$ and $\tilde{g}=1$. The minus sign indicates underestimation. \label{tab: layer unit sigma}}
\newcolumntype{C}{>{\centering\arraybackslash}X}
\begin{tabularx}{0.5\columnwidth}{cc|CCC}
    \hline\hline
    \multirow{2}{*}{Relative error of $\sigma$ (\%)} & & \multicolumn{3}{c}{Number of units per layer}
    \\
    & & 64 & 128 & 256
    \\
    \hline
    \multirow{3}{*}{Number of hidden layers} 
    & 1 & 4.1 & -2.1 & 4.2
    \\
     & 2 & 0.68 & 0.36 & 0.62
    \\
     & 3 & 0.16 & 0.26 & 0.12
    \\
    \hline\hline
\end{tabularx}
\end{table}
Finally, we explore how the NN's size influences our results. Tables \ref{tab: layer unit gamma} and \ref{tab: layer unit sigma} detail the errors in measuring $\gamma(l_{\mathrm{end}},0)$ and $\sigma(l_{\mathrm{end}},0)$, respectively, each considering different numbers of hidden layers and units per layer, with settings of $N=1$ and $\tilde{g}=1$. From these, it's clear that more hidden layers lead to better accuracy. However, while this improvement tends to level off for $\gamma(l_{\mathrm{end}},0)$, the accuracy for $\sigma(l_{\mathrm{end}},0)$ continues to benefit from additional layers. This difference might be unique to the architecture we've used. We observed similar trends under other settings for $N$ and $\tilde{g}$.

\end{widetext}

\bibliography{main}

\end{document}